\documentclass[twocolumn,aps,pre,amsmath,showpacs,tufte-book]{revtex4-1}
\usepackage{graphicx,amsmath,leftidx,amsfonts}
\usepackage{subfig,array,booktabs,diagbox}
\usepackage[normalem]{ulem}
\usepackage{dcolumn,bm,amssymb,indentfirst}
\usepackage{color,soul}
\setstcolor{green}
\usepackage{caption}
\begin{document}
\author{Bingyu Cui$^{1}$, Jonathan F. Gebbia$^{3}$, Josep-Lluis Tamarit$^{3}$,
Alessio Zaccone$^{1,2}$}
\affiliation{${}^1$Statistical Physics Group, Department of Chemical
Engineering and Biotechnology, University of Cambridge, New Museums Site, CB2
3RA Cambridge, U.K.}
\affiliation{${}^2$Cavendish Laboratory, University of Cambridge, JJ Thomson
Avenue, CB3 0HE Cambridge,
U.K.}
\affiliation{${}^3$Grup de Caracterizacio de Materials, Departament de Fisica,
EEBE and \\ Barcelona Research Center in Multiscale Science and Engineering, \\
Universitat Politecnica de Catalunya, Eduard Maristany, 10-14, 08019 Barcelona,
Catalonia}
\begin{abstract}

We use a microscopically motivated Generalized Langevin Equation (GLE) approach
to link the vibrational density of states (VDOS) to the dielectric response of
orientational glasses (OGs). The dielectric function calculated based on the
GLE is compared with experimental data for the paradigmatic case of two OGs:
Freon 112 and Freon 113, around and just above $T_g$. The memory function is related
to the integral of the VDOS times a spectral coupling function
$\gamma(\omega_p)$, which tells the degree of dynamical coupling between
molecular degrees of freedom at different eigenfrequencies.
The comparative analysis of the two Freons reveals that the appearance of a
secondary $\beta$ relaxation in Freon 112 is due to cooperative dynamical
coupling in the regime of mesoscopic motions caused by stronger anharmonicity
(absent in Freon 113), and is associated with comparatively lower boson peak in
the VDOS. The proposed framework brings together all the key aspects of glassy
physics (VDOS with boson peak, dynamical heterogeneity,
dissipation, anharmonicity) into a single model.
\end{abstract}

\title{Disentangling $\alpha$ and $\beta$ relaxation in orientationally
disordered crystals with theory and experiments}
\maketitle

\section{Introduction}
Structural glass (SG) formers, which are usually obtained from supercooled
liquids in which translational and orientational degrees of freedom are frozen
below the glass transition temperature $T_g$, exhibit a complex response
function on vibrational excitations
\cite{DEAN1972,Donth,Martinez2001,CAAngell2000}.
When they undergo a rapid cooling to avoid crystallization, some
anomalous physical properties emerge. For example, as temperature
decreases, the relaxation time generally shows a stronger increase, faster than
what is
given by the Arrhenius law (super-Arrhenius behavior). For such cases, the
temperature ($T$) dependence of relaxation time ($\tau$) is given through the
empirical Vogel-Fulcher-Tammann (VFT) law \cite{KLNgai2000JCP,Fstickel1995}, or
by physically-motivated double-exponential dependence of $\tau$ on $T$, which
includes the dependence on the steepness of interatomic repulsion and on
thermal expansion via the more recent Krausser-Samwer-Zaccone (KSZ)
relation~\cite{Krausser}. To account for the deviation of the Arrhenius
temperature dependence, i.e., for the faster increase of the relaxation time
(or viscosity), the kinetic fragility index is defined as $m = \frac{\partial
\log{\tau}}{\partial (T_g/T)}\big |_{T = T_g}$, ranging between $\approx 16$
(strong glasses) and $200$ (fragile glasses) \cite{Rbohmer1993,ANGELL1988}.

In addition to the structural glasses, orientational glasses (OGs) can be
obtained from orientationally disordered (OD) phases or plastic phases
\cite{Drozd2006,RBrand2002,Ramos1997}. OD phases are crystal
lattices in
which weakly interacting molecules are orientationally disordered. On cooling,
some OD phases exhibit the same features as structural (canonical) glass
formers \cite{RBrand2002, GAVdovichenko2015, IVSharapova2010, Pardo2006}. With
respect to the fragility index, OGs are ususally strong \cite{Rbohmer1993}
whereas for SGs a wide range of fragility values are found, the most fragile
being the cis- or trans-decahydronaphthalene ($m\approx147$)
\cite{Kalyan2002}. The most fragile OGs known to date contains Freon 112
(CCl$_2$F-CCl$_2$F, hereinafter F112) with $m=68$ \cite{Pardo2006}. On the
other hand, for Freon 113 (CCl$_2$F-CClF$_2$, hereinafter F113), the kinetic
fragility index was calculated to be $m=127$ , which is the highest so far
reported for an OG \cite{Vispa2017}.

In an effort to explain the various glassy anomalies and dynamical behaviour, 
mode-coupling theory provides, among other predictions, a good description of dielectric
relaxation of liquids for temperatures higher than the liquid crossover
temperature $T_{c}$~\cite{Blochowitz}. In spite of this, the main relaxation mechanism by which
supercooled liquids undergo a liquid-solid transition-like at or around $T_g$
has remained elusive \cite{Albert1308}. The $\alpha$-relaxation, typically
associated with collective and strong cooperative motions of a large number of
entities rearranging in a long-range correlated way, is related to the slowest
decay of density correlations and is widely observed in dielectric and
mechanical responses.

For supercooled liquids, the empirical Kohlrausch stretched-exponential
function $\sim\exp{(-(t/\tau)^\beta)}$
provides a good empirical fit for the dielectric loss of the $\alpha$-relaxation, upon
taking the Fourier transform from time into frequency domain. Further, starting
from the first principle assumption that the
microscopic Hamiltonian can be modelled using a classical particle-bath
coupling of the
Caldeira-Leggett type, a simple and explicit relation between the
dielectric relaxation function and the VDOS of SGs has been
presented to provide a good interpretation of the $\alpha$-peak and
stretched-exponential relaxation, through a memory function of
friction~\cite{CuiPRE2017}.

In addition to $\alpha$-relaxation, an extra shoulder or wing also decorates the
imaginary part of the dielectric response, which is referred to as the
$\beta$-relaxation, or as Johari-Goldstein or secondary relaxation.
As discovered by Johari and Goldstein \cite{JohariTJCP1970} in glasses of rigid
molecules and as described by the Ngai coupling model \cite{NGaiPRE1998}, the
secondary relaxation involves the motion of the entire molecule. Knowing the
underlying mechanism of $\beta$-relaxation, is
of great importance for understanding many crucial unresolved issues in glassy
physics and materials science and consequently for a wide potential
application in technologies, ranging from glass transitions, deformation
mechanisms, to diffusion and the breakdown of the Stokes-Einstein relations,
physical ageing, as well as the conductivity of ionic liquids and the stability
of glassy pharmaceuticals and biomaterials. Yet, the
nature and mechanism of the $\beta$-relaxations are still not clear
\cite{PLunk2000, YU20012, NGAI2000, NGai1998}.

In order to understand the puzzling origin of the $\beta$ relaxation it is instructive
to consider systems with very similar molecular structure and yet exhibiting widely different
relaxation behaviour. Such systems can be found in the realm of OGs.

Previous study on thermal conductivities of Freon 112 and Freon 113 (F113) reveals the
existence of quasilocalized low-energy vibrational modes (soft harmonic
oscillators as described through the soft potential model~\cite{Parshin1994})
at energy lower than the values of the maximum of the boson peak compared with
other OGs, which results in an increase of the VDOS \cite{GAVdovichenko2015}. It
was thought that the high values of kinetic $(m=127)$
fragility of F113 is produced by strong orientational correlations, which
is evidenced by low values of the stretching exponent in Kohlrausch
stretched-exponential function close to $T_g$, where only $\alpha$-relaxation
is observed with no sign of the $\beta$-relaxation. On the contrary, in
dielectric spectra of Freon 112 (F112), $\beta$-relaxation emerges as temperature decreases
to $T_g$ and becomes evident below $T_g$.

The above experimental facts are the origin of our interest in applying a
microscopic theoretical model to plastic crystals. In particular, freons F112
and F113 are chemically and molecularly similar compounds displaying glassy
states (they both belong to the series C$_2$X$_{(6-n)}$Y$_n$, with X, Y$ = $Cl,
F, Br, and $n=0,...,6$.), but with completely different dynamics and
relaxation. This provides  a unique opportunity to explain, from a microscopic
point of view, the physical origin of secondary $\beta$ relaxation.

We therefore developed a modified theoretical model in the spirit of
Ref.~\cite{CuiPRE2017}, to account for both $\alpha$ and $\beta$ relaxation and
we apply it to OG states of freons F112 and F113. From the analysis of
experimental data, it is evident that:
i)      the proposed generalized Langevin equation (GLE) theory successfully
describes both $\alpha$ and $\beta$ relaxation process in the dielectric
response, by using the experimentally measured VDOS as input;
ii)     the model provides a new insight into the dynamical origin of the
secondary relaxation;
iii) the model also clarifies, for the first time, which eigenmodes dynamically
couple with the secondary relaxation process.
This framework represents a new microscopic modelling of the glassy relaxation
in orientationally disordered crystals for which no theoretical model was
available so far.


\section{Theory}
Focusing on a tagged particle (e.g. a molecular subunit carrying a partial
charge which reorients under the electric field), it is possible to describe
its motion under the applied field using a particle-bath Hamiltonian of the
Caldeira-Leggett type, in the classical dynamics regime~\cite{CuiPRE2017}.
The particle's Hamiltonian is bi-linearly coupled to a bath of harmonic
oscillators which represent all other molecular degrees of freedom in the
system~\cite{Zwanzig1973}. Any complex system of oscillators can be reduced to
a set of independent oscillators by performing a suitable normal mode
decomposition. This allows us to identify the spectrum of eigenfrequencies of
the system, i.e. the experimental VDOS, with the spectrum of the  set of
harmonic oscillators forming the bath.

\subsection{Particle-bath Hamiltonian and GLE}
The particle-bath Hamiltonian under a uniform AC electric field, is given
by \cite{CuiPRE2017}: $H=H_P+H_B$ where $H_P=P^2/2m+V(Q)-q_e Q E_0\sin{(\omega
t)} $ is the Hamiltonian of the tagged particle with the
external electric field ($q_e$ is the charge carried by the particle),
$H_B=\frac{1}{2}\sum_{\alpha=1}^N\left[\frac{P_{\alpha}^2}{m_{\alpha}}+m_{\alpha}\omega_{\alpha}^2\left(X_{\alpha}
-\frac{F_{\alpha}(Q)}{\omega_{\alpha}^2}\right)^{2}\right]$ is the Hamiltonian
of the bath of harmonic oscillators that are coupled to the tagged
particle~\cite{Zwanzig1973}.
Two parts in $H_B$ are of physical interest: The first part is the ordinary
harmonic
oscillator; the second is the coupling term between the tagged particle
position $Q$ and the
bath oscillator position $X_{\alpha}$. The coupling function is taken to be
linear in the displacement of the particle,
$F_{\alpha}(Q)=c_{\alpha}Q$, where $c_{\alpha}$ is known as the strength of
coupling between the tagged atom and the $\alpha$-th bath oscillator.
Hence, there is a spectrum of coupling constants $c_{\alpha}$ by which each
particle interacts with all other molecular degrees of freedom in the
system. This spectrum of coupling strengths will play a major role in the
subsequent analysis. The equation of motion for the tagged particle can then be
derived straightforwardly, which leads to the following GLE
\begin{equation}
\ddot{q}=-V'(q)-\int_{-\infty}^t \nu(t')\frac{dq}{dt'}dt'+q_eE_0\sin{(\omega
t)}.
\end{equation}
where the non-Markovian friction or memory kernel $\nu(t)$ is given by:
\begin{equation}
\nu(t)=\sum_\alpha\frac{c_\alpha^2m_\alpha}{\omega_\alpha^2m}\cos{(\omega_\alpha
t)}.
\end{equation}
Note we have converted into rescaled coordinates for standard normal-mode
analysis: $q=Q\sqrt{m}$. This means $V(Q)$ and $V(q)$ are basically different
functions.  We have also redefined $q_e=e/\sqrt{m}$ as the (partial) reduced
charge in rescaled coordinates.
Then we can let the spectrum be continuous and $c_{\alpha}$ be a function of
eigenfrequency $\omega_p$, which leads to the following expression for the friction kernel:
\begin{equation}\label{eq:nu}
\nu(t)=\int_0^{\infty}d\omega_p
D(\omega_p)\frac{\gamma(\omega_p)^2}{\omega_p^2}\cos{(\omega_p t)},
\end{equation}
where $\gamma(\omega_p)$ is the continuous spectrum of coupling constants, i.e.
the continuous version of the discrete set $\{c_{\alpha}\}$ averaged over all
tagged particles.

For any given (well behaved) VDOS function $D(\omega_{p})$, the existence of a
well-behaved function $\gamma(\omega_p)$ that satisfies Eq. (3) is guaranteed
by the fact that we can always decompose $\nu(t)$ into a basis of
$\{\cos{(\omega_p t)}\}$ functions, by taking a cosine transform. The inverse
cosine transform in turn gives the spectrum of coupling constants
$\gamma(\omega_p)$ as a function of the memory kernel:
\begin{equation}\label{eq:gamma}
\gamma^2(\omega_p)=\frac{2\omega_p^2}{\pi
D(\omega_p)}\int_0^{\infty}\nu(t)\cos{(\omega_pt)}dt.
\end{equation}
This coupling function contains information on how strongly the particle's
motion is coupled to the motion of other particles in a mode with vibrational
frequency $\omega_p$. This is an important information, because it tells us
about the degree of long-range anharmonic couplings in the motion of the
molecules.

\subsection{Memory function modelling}
Looking at Eq. (\ref{eq:nu}), it is evident that the particle-bath Hamiltonian
does not
provide any prescription to the form of the memory function $\nu(t)$, which can
take any form depending on the values of the coefficients
$c_{\alpha}$~\cite{Zwanzig1973}.
Hence, a shortcoming of particle-bath models is that the functional
form of $\nu(t)$ cannot be derived a priori for a given system, because, while
the VDOS is certainly an easily accessible quantity from simulations of a
physical system, the spectrum of coupling constants $\{c_{\alpha}\}$ is
basically a phenomenological parameter.

However, for a supercooled liquid, the time-dependent friction, which is
dominated by slow collective dynamics, has been famously derived within kinetic
theory (Boltzmann equation) using a mode-coupling type approximation by
Sjoegren and Sjoelander ~\cite{LSjogren1979}, and
is given by the following elegant expression:
\begin{equation}
\nu(t)=\frac{\rho k_{B}T}{6\pi^2 m}\int_{0}^{\infty}dk k^{4}
F_{s}(k,t)[c(k)]^{2} F(k,t)
\end{equation}
where $c(k)$ is the direct correlation function of
liquid-state theory,
$F_{s}(k,t)$ is the self-part of the intermediate scattering function (ISF)
$F(k,t)$~\cite{LSjogren1979}. All of these quantities are functions of the
wave-vector $k$. Clearly, the integral over $k$ leaves a time-dependence of
$\nu(t)$ which is controlled by the product $F_{s}(k,t)F(k,t)$. For a
chemically homogeneous system,
$F_{s}(k,t)F(k,t)\sim F(k,t)^{2}$, especially in the long-time regime.

From theory and simulations, we know that in supercooled liquids $F(k,t)\sim
\exp{(-(t/\tau)^b)}$ for some $\tau$ and $b$, when only $\alpha$-relaxation is
present.
When both $\alpha$ and $\beta$ relaxation are present, the ISF has a two-step
decay (one for $\alpha$ and one for $\beta$)~\cite{Donth}.
It is easy to check that a two-step decay of the ISF within Eq. (5) is perfectly
compatible with a memory function $\nu(t)$ given by a sum of two
stretched-exponential terms.

While the elegant relation by Sjoegren and Sjoelander Eq. (5) relies on
mode-coupling type assumptions which may be questionable below $T_g$, we also
point out that a more physically meaningful justification comes from its
ability to generate an ISF $F(k,t)$ with a two-step
decay in time for F112 upon taking $\nu(t)$ as a sum of two stretched-exponentials, which
is also compatible with the dielectric data (see fittings below). This qualitative behavior for
the ISF with a two-step decay has been demonstrated for the Freon 112 system in
simulations, e.g. Ref.~\cite{Affouard_PRE}, and also in experiments
~\cite{Affouard_JCP}. Hence, despite the fact that the Sjoegren and Sjoelander
relation relies on assumptions of mode-coupling type, the relationship between
our memory function and the intermediate scattering function is physically
meaningful and supported by data in the literature.

Hence, in light of the above discussion, we will take the following
phenomenological expression for our memory function
\begin{equation}
\nu(t)=\nu_0\sum_ie^{-(t/\tau_i)^{b_i}},
\end{equation}
where $\tau_i$ is a characteristic time-scale, with $i=1$ for pure $\alpha$
relaxation and $i=1,2$ for co-existing $\alpha$ and $\beta$ relaxation.
$\nu_{0}$ is a constant
pre-factor.


\subsection{Dielectric response and link with the VDOS}
Following  the same steps as those described in Ref.\cite{CuiPRE2017}, upon taking the GLE Eq. (1) as the starting point, we
obtain the complex dielectric function as
\begin{equation}\label{eq:epsomega}
\epsilon^*(\omega)=1-A\int_{0}^{\omega_{D}}\frac{D(\omega_{p})}{\omega^2-i\omega\tilde{\nu}(\omega)-\omega_p^2}
d\omega_{p}
\end{equation}
where $A$ is an arbitrary positive rescaling constant, $\omega_{D}$ is the
Debye cut-off frequency (i.e.
the highest eigenfrequency in the VDOS spectrum), and tilde over $\nu$ denotes
Fourier transformation. As one can easily verify, if
$D(\omega_{p})$ were given by a Dirac delta, one would recover the standard
simple-exponential (Debye) relaxation.

\begin{figure}
\centering
\includegraphics[height=6cm ,width=8cm]{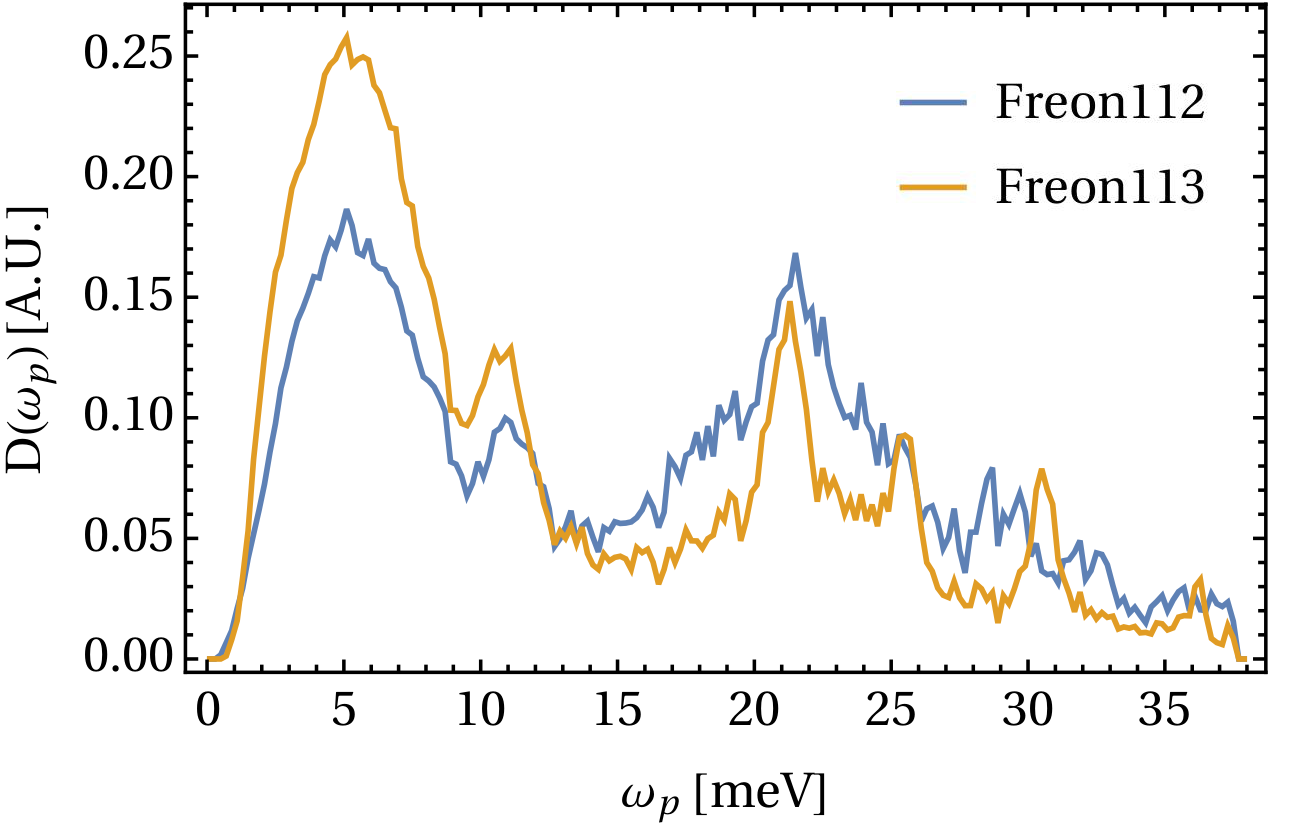}
\caption{Experimental vibrational density of states (VDOS) for Freon 112 (blue)
and Freon 113 (yellow). The data for Freon 112 were published in
Ref.~\cite{IVSharapova2010}, while the data for Freon 113 were taken from
Ref.~\cite{Vispa2017}.}
\label{fig:FreonDOS}
\end{figure}

The VDOS is an important key input to the theoretical framework. The essential
experimental VDOS were measured by means of inelastic neutron scattering using
the direct spectrometer MARI of the ISIS facility (UK) and are shown in Fig.
\ref{fig:FreonDOS}. The VDOS of F113 clearly exhibits a much more significant
excess of low-frequency (boson-peak) modes, with respect to F112, in the range
$2-5$meV.

For F113, we use only one stretched-exponential term in the memory function $\nu(t)$,
hence $i=1$ in Eq. (6).
For F112, instead, $\nu(t)$ is the sum of two terms ($i=1,2$ in Eq. (6)), both
of which are
stretched-exponential. The first term
represents mainly the $\alpha$ process although it also affects the $\beta$
relaxation (hence the two are coupled, as one can anticipate in the spirit of
the Ngai coupling model~\cite{NGAI2000}). The
second term describes only $\beta$ relaxation. Thus, the time-scale of
$\beta$ relaxation is not identically equal to the time-scale of the second
stretched-exponential parameter, which is $\tau_2$. This amounts to the fact
that $\beta$ relaxation is a process which is cooperative (hence coupled to the
$\alpha$) and at the same time
quasi-localized.

In terms of physical meaning, $\tau_1$ represents the time-scale of
$\alpha$-relaxation, and the stretching exponent is related to the distribution
of escape times from larger metastable basins in the glassy energy landscape.
This is because stretched-exponential form arises from the integral average of
simple exponential decays weighted by a distribution of time-scales, the
broader the distribution the lower the resulting
stretching-exponent~\cite{Johnston}. Similarly, the second
stretched-exponential required to describe $\beta$-relaxation is possibly
related to the distribution of smaller wells within the same meta-basin.

\section{Comparison with experimental data of dielectric loss}
\begin{figure}[tb]
\centering
{%
\includegraphics[height=5.5cm ,width=8cm]{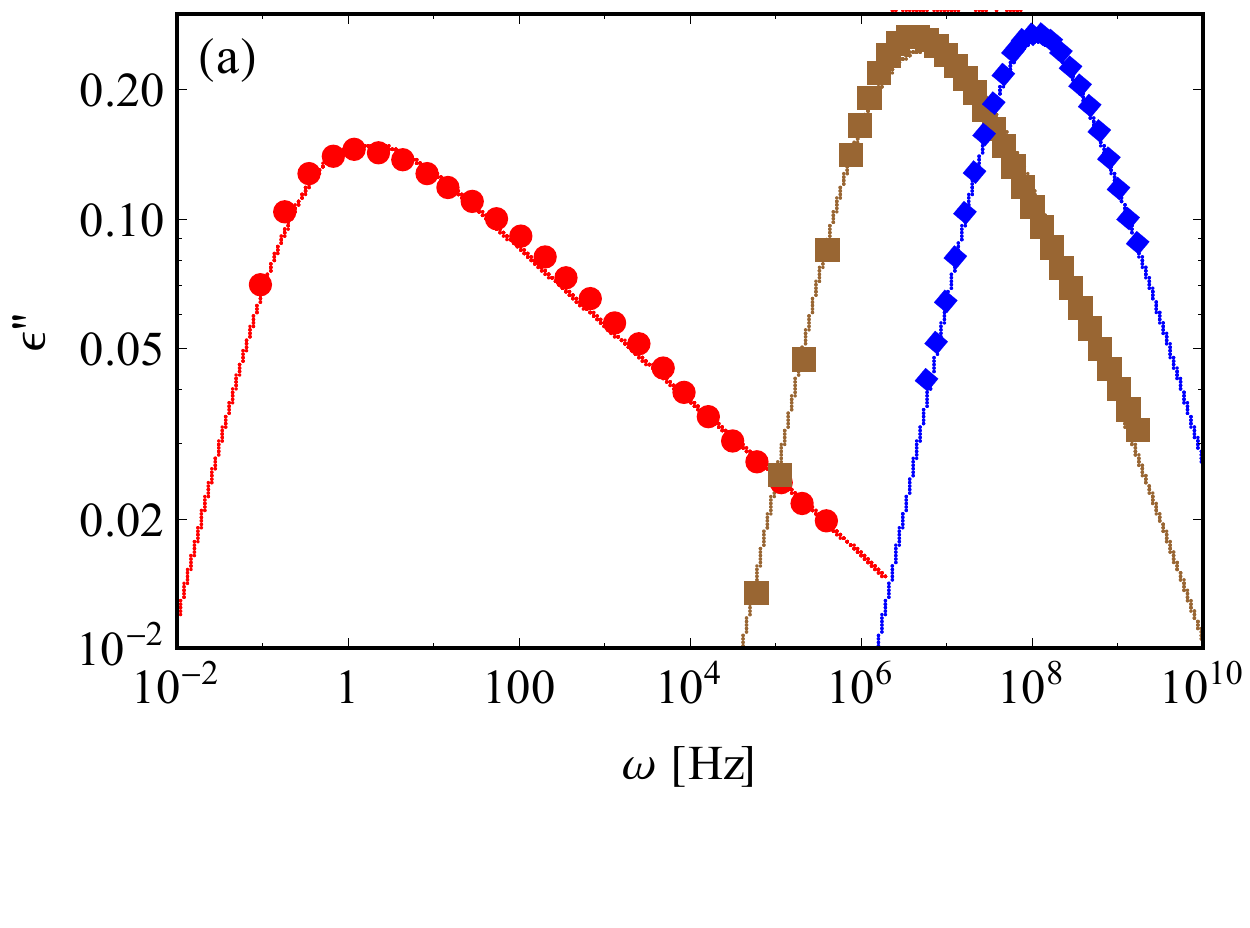}\label{Freon112}}\hfill
{%
\includegraphics[height=5.5cm ,width=8cm]{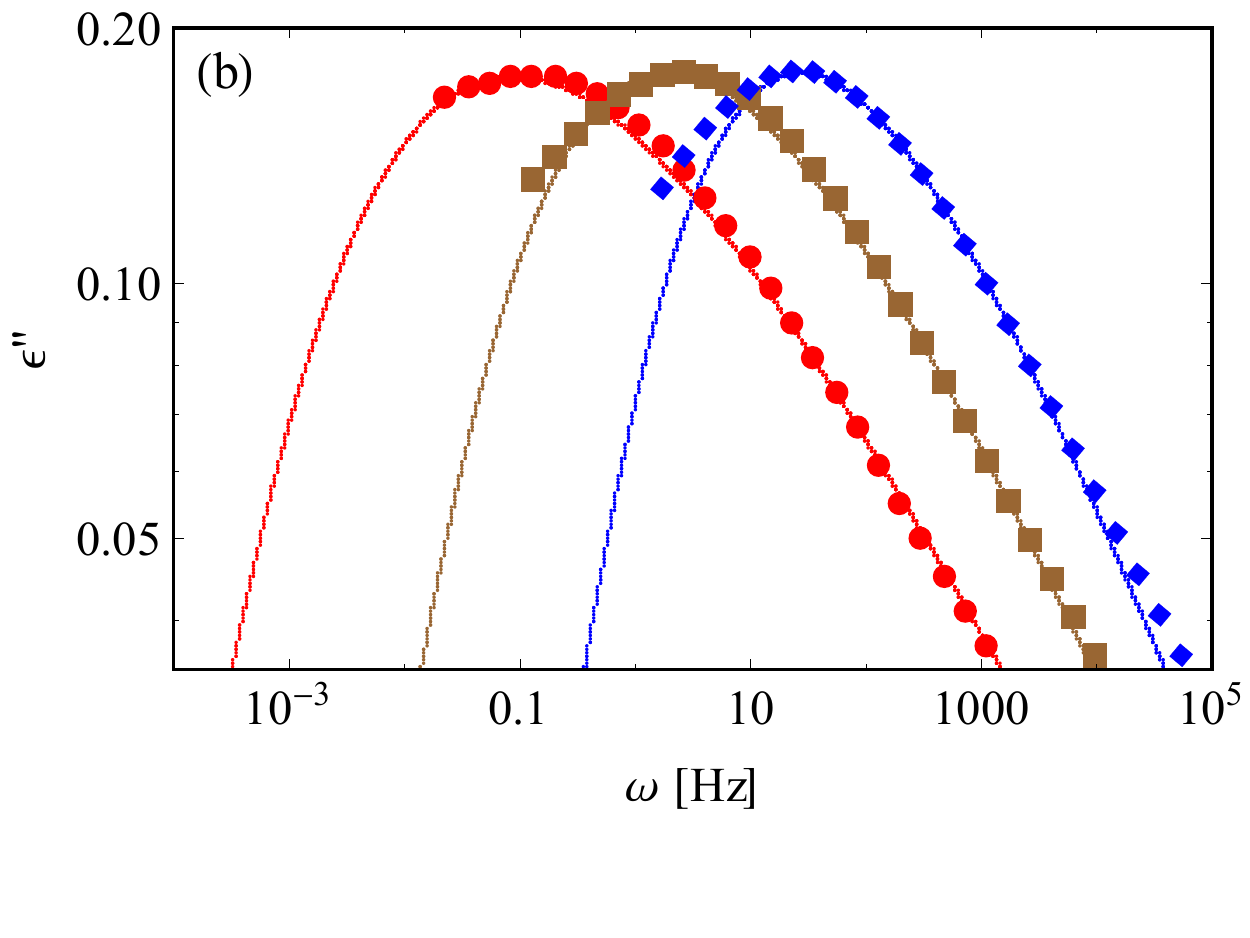}\label{Freon113}}
\label{Freons}
\caption{Fitting of experimental data using the proposed theoretical model for
Freon 112 (a) at 91 K (red circles), 115 K  (brown squares) and 131 K (blue
diamonds) and for Freon 113 (b) at 72 K (red circles), 74 K (brown squares) and
76 K (blue diamonds). Solid lines are the theoretical model presented here. A
rescaling constant was used to adjust the height of the curves since the data
are in arbitrary units. Experimental data for Freon 112 were taken from
Ref.~\cite{Pardo2006}, while data for Freon 113 were taken from
Ref.~\cite{Vispa2017}.
}
\label{fig:epsF112&F113}
\end{figure}

Fitting parameters for F112 and F113 at
different temperatures are listed in Table I \& II and resulting fittings of
dielectric loss are displayed in Fig. \ref{fig:epsF112&F113}.
\begin{table}[tbp]
\centering
\begin{tabular}{lccc}
\hline
Temperature &91K &115K &131K\\ \hline
$b_1$ &0.45 &0.625 &0.7\\
$\tau_1$ (seconds) &0.558 &$3.12\cdot10^{-7}$ &$6.99\cdot10^{-9}$\\
$b_2$ &0.2 &0.56 & \\
$\tau_2$ (seconds) &$1.55\cdot10^{-2}$ &$5.48\cdot10^{-8}$ & \\
$\nu_0$ &$4\cdot10^6$ &$3.9\cdot10^6$  &$6.3\cdot10^6$\\ \hline
\end{tabular}
\caption{Parameters of the memory function for Freon 112.}
\begin{tabular}{lccc}
\hline
Temperature &72K &74K &76K\\ \hline
$b$ &0.26 &0.3 &0.35\\
$\tau$ (seconds) &7.133 &0.326 &$2.38\cdot10^{-2}$\\
$\nu_0$ &$8.28\cdot10^6$ &$7.28\cdot10^6$  &$6.28\cdot10^6$\\ \hline
\end{tabular}
\caption{Parameters of the memory function for Freon 113.}
\end{table}

For the fitting procedure, we have assumed that $D(\omega_p)$ and the overall
scaling for the height of curve, $A$, are $T$-independent.
\begin{figure}[tb]
\centering
{%
\includegraphics[height=5.5cm ,width=8cm]{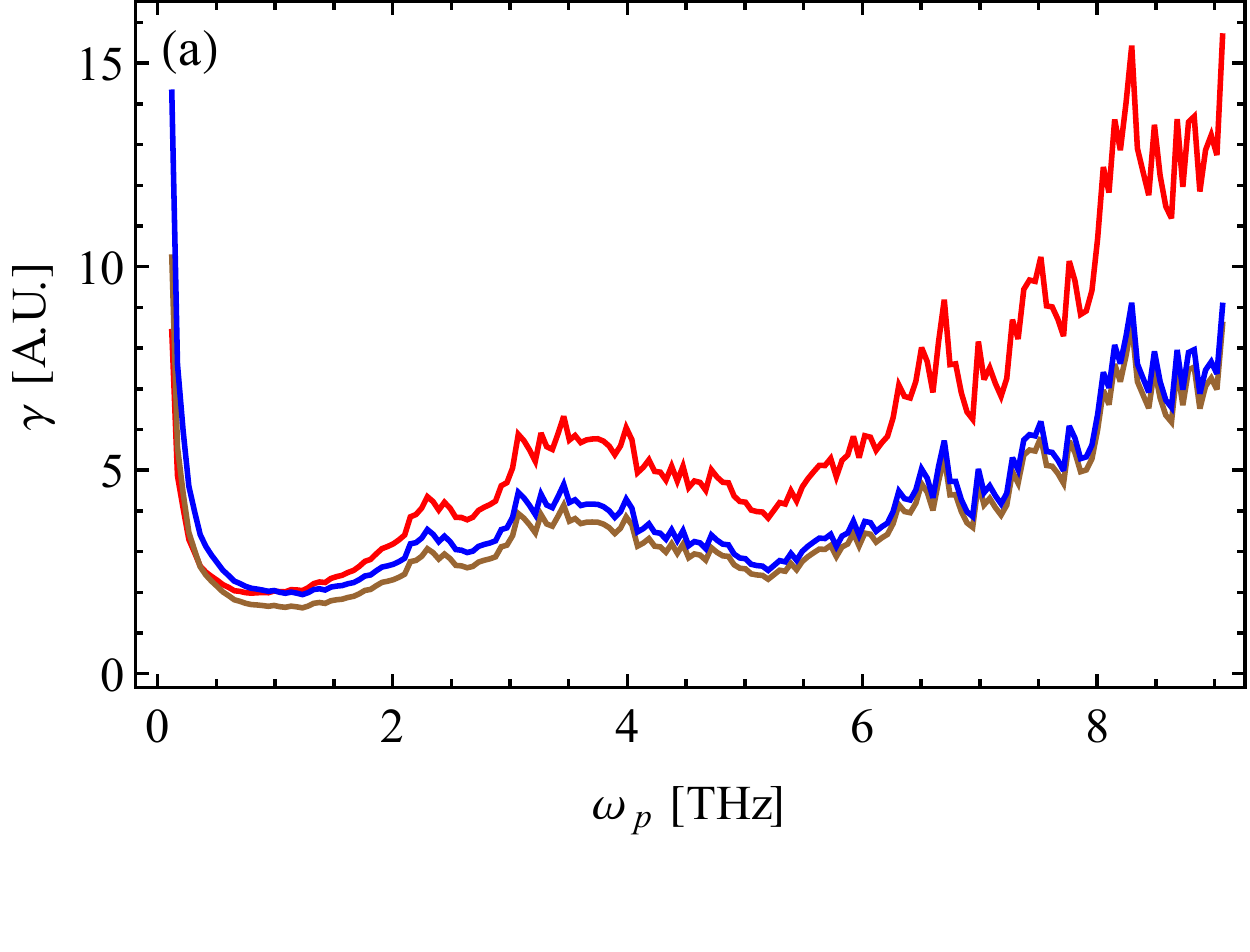}
\label{coupling112}}\hfill
{%
\includegraphics[height=5.5cm ,width=8cm]{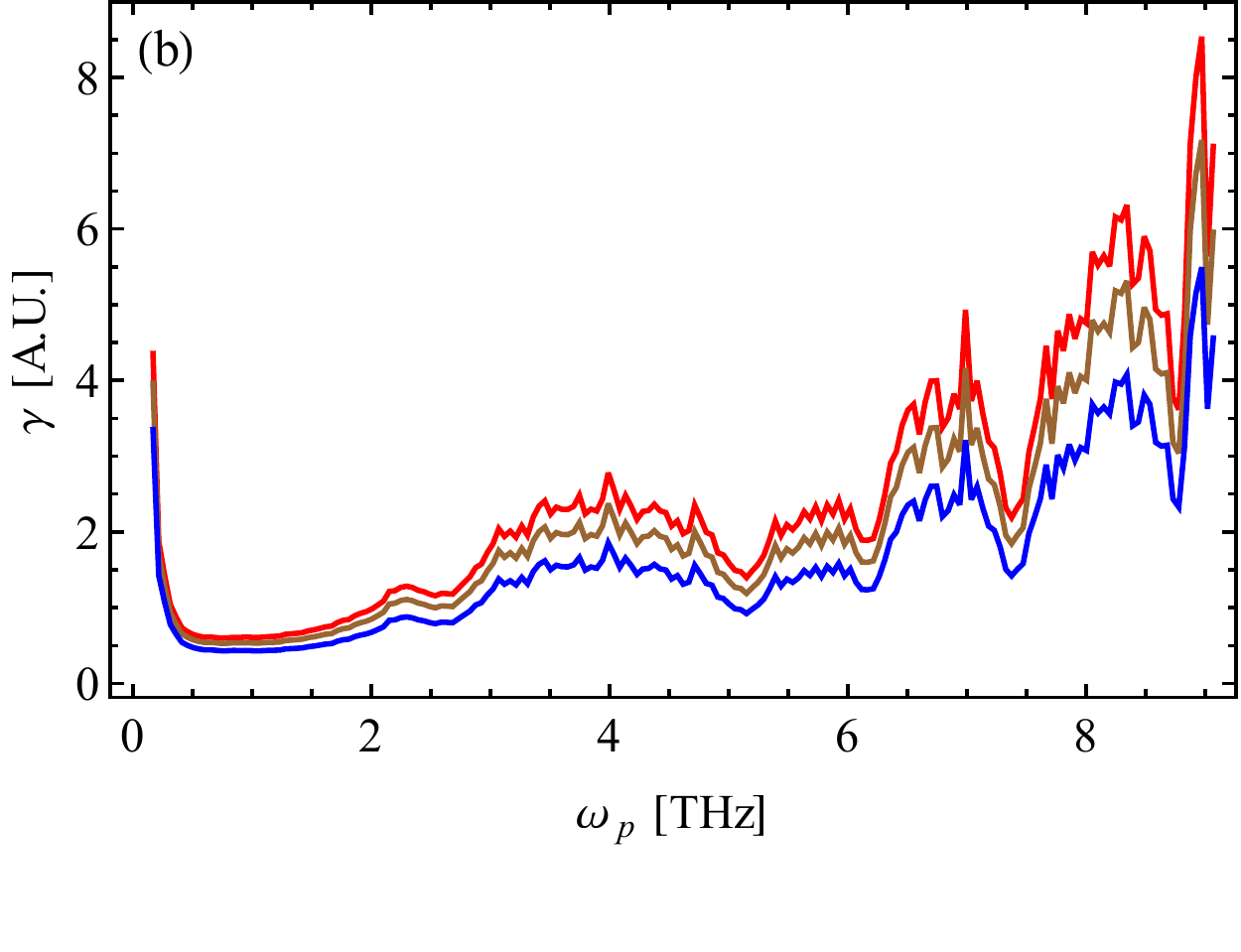}\label{coupling113}}
\label{coupling}
\caption{
Spectrum of coupling constants of Freon 112 (a) and Freon 113 (b) as a function
of the vibrational eigenfrequency computed according to Eq. (\ref{eq:gamma})
using the phenomenological memory functions $\nu(t)$ used in the fitting of
dielectric response in Fig. \ref{fig:epsF112&F113}, with same colour setting
for the different temperatures.}
\label{fig:gamma}
\end{figure}

\section{Physical mechanism of secondary relaxation}
To physically understand the difference between F112 and F113, their dynamical
coupling functions (Eq. (\ref{eq:gamma})) have been analysed (see Fig.
\ref{fig:gamma}). In general, the coupling spectrum decays from the highest
Debye cut-off frequency of short-range high-frequency in-cage motions, down to
the low eigenfrequency part where the coupling goes up with decreasing
$\omega_p$ towards zero, due to phonons or phonon-like excitations, which are
collective and long-wavelength and hence result in a larger value of $\gamma$.


There is a substantial difference between F112 and F113, especially in the
middle part of the
coupling spectrum where F112 shows much stronger coupling, which corresponds to
medium-range correlated motions. This means that motions are strongly coupled
also in the
intermediate eigenfrequency domain, where modes are typically quasi-localized,
which corresponds to mesoscopic string-like
motions~\cite{Douglas} typically associated with
$\beta$-relaxation~\cite{Samwer}. In addition, the F113 spectrum is overall
comparatively much lower in that energy regime, which clearly indicates that,
for F113, the intermediate part of the coupling spectrum, i.e. the one of
mesoscopic and string-like motions, is scarcely populated and one has a steep
decay from the short-range high-frequency in-cage motions to the
long-wavelength phonon-like excitations, with not much in between in the
mesoscopic range.
Hence in F113, the anharmonicity is much less prominent and intermediate
excitations are not important. This origin of the secondary relaxation aligns
with the simulation results of Refs.~\cite{Harrowell,Procaccia} which point at
the cooperative, though localized or quasi-localized, nature of secondary
relaxation.

This also gives insights into the difference in the form
of the memory function used for the fittings of the two Freons. Upon
focusing on the integration in Eq. (\ref{eq:gamma}): the integral of $\nu(t)$
from $0$ to
$\infty$ increases from high $\omega_p$ (short-range and fast vibration) to
low $\omega_p$ (long-range and slow vibration), since for slow collective
vibration there is clearly much more extended friction due to contact between
many
particles all moving at the same time. Thus, the integral factor definitely
contributes to the coupling being overall stronger for F112 than for F113.
However, also the boson peak contributes to the coupling of F112 being larger
than that of F113 (via the VDOS in the denominator of Eq. (\ref{eq:gamma})) in
the specific frequency range that corresponds to the boson
peak). The boson peak maximum (in $D(\omega)/\omega^2$, not shown) for both
materials is of the order of $2-3$ meV, i.e., $\approx0.5-0.7$ THz, which
corresponds virtually to the lowest minimum in the coupling function (see Fig.
\ref{eq:gamma}) where, in addition, the minimum value is much lower for F113
(with larger boson peak) than for F112.
That means that in such region not only we have larger dynamical
coupling for F112 due to stronger medium-range correlations/anahrmonicity, in
general, but also for the additional effect of boson peak (soft weakly-coupled
modes, see Fig. 1) being smaller for F112
in that regime of vibrations.

As far as temperature effects on the coupling strength, we must point out first
that, due to the fragility difference between the two freons, the temperature
range in which fittings were performed are noticeably different. For F113 ($T_g
= 71$ K) experimental dielectric functions are available at the highest reduced
temperature of $T_r=76/71=1.07$, whereas for F112 ($T_g=90$ K) the highest
value is around $T_r=131/90=1.46$. Bearing this in mind, it can be noticed that
upon increasing temperature, the "going up" tail at decreasing $\omega_p$
towards zero becomes smaller,
which means less phonon-like modes. In general, absolute coupling values shift
down (lower coupling) with the increase of temperature, as expected, and the
decay of correlated motions from high $\omega_p$ to low $\omega_p$ becomes also
somewhat steeper with increasing $T$.


\section{Discussion and conclusions}
The stronger coupling between collective and individual motions for F112 could
be a physical explanation of why in the dielectric study of F112
\cite{Pardo2006} the authors described so many problems to disclose $\alpha$-
from $\beta$- relaxation. For F112 collective vibrations, medium-ranged and
slow motions are much more important than for F113, in such a way that
individual molecular motions ($\beta$-relaxation) should correlate, i.e. are
much more coupled, with motions of surrounding molecules (collective motions
associated with the $\alpha$-relaxation). And, even more, if slow vibrations
are more important and more heterogeneous in F112, this should mean stronger
coupling between collective and individual motions, so then, much more phonon
scattering for F112 and, as a consequence, lower thermal conductivity for F112
than for F113, as it has been experimentally shown (see Fig. 5 in
\cite{GAVdovichenko2015}). In addition, it should be emphasized that the higher
thermal conductivity for F113, analysed in terms of the soft-potential model,
was also attributed to the low coupling strength between sound waves and the
soft quasi-localized modes.
Moreover, the dynamical coupling function $\gamma$ extends over a frequency
range much broader than that of the boson peak, and thus the role of the boson
peak is confined to a specific frequency range which is around the minimum in
the coupling spectrum. The fact that boson peak is stronger for F113 leads to a
lower coupling in that region and contributes to the already lower coupling of
F113 compared to F112 in that region. Because the boson peak is associated with
soft modes, which "break" the coherence of phonons (hence more phonon
scattering), it leads to even lower coupling in the boson peak frequency range
for F113.\\

In conclusion, we have presented a new approach which makes it possible to
directly link the vibrational density of states of orientational glasses
measured experimentally with the macroscopic dielectric response and the
underlying heterogeneous dynamics. Furthermore, the model effectively accounts
also for the medium- and long-range anharmonic coupling among molecular degrees
of freedom and allows one to disentangle $\alpha$ and $\beta$ relaxation on the
basis of the extent of dynamical coupling in different eigenfrequency sectors
of the vibrational spectrum. The appearance of secondary $\beta$ relaxation is
associated with higher values of the dynamical coupling strength of correlated
particle motions in the regime of mesoscopic quasi-localized motions (e.g.
string-like motions, vortices, etc. \cite{Samwer}) and is also promoted by a
lower excess of soft modes in the boson peak frequency range.

In our model, we require two forms of stretched exponentials in the memory
function, hence two relaxation times, to fit both alpha and $\beta$ relaxations.
The $\beta$-relaxation process cannot be recovered with only one stretched
exponential (i.e. with only one term in the memory function). One of the
stretched exponentials dominates the $\alpha$ peak while the co-existing effect of
two stretched exponential terms in the memory function gives rise to the
$\beta$ or secondary relaxation. In other words, the two terms of memory function both
affect the secondary relaxation, whereas only one of them controls the $\alpha$
relaxation. This implies that there is indeed a deep microscopic dynamical
coupling between the two relaxation processes, which has not been unveiled so
far. In future work this framework will be used to provide more microscopic
insights into the dynamical nature of this coupling and in the context of the
Ngai coupling model~\cite{NGai1998}.

\begin{acknowledgements}
B. C acknowledges the financial support of CSC-Cambridge Scholarship. J.Ll. T.
acknowledges MINECO (FIS2017-82625-P) and the Catalan government (SGR2017-042)
for financial support.
\end{acknowledgements}

\end{document}